# Fermi-liquid nonadiabatic highly-compressed cesium iodide superconductor

Evgeny F. Talantsev[1,2]

[1]M.N. Mikheev Institute of Metal Physics, Ural Branch, Russian Academy of Sciences, 18, S. Kovalevskoy St, Ekaterinburg, 620108, Russia
[2]NANOTECH Centre, Ural Federal University, 19 Mira St, Ekaterinburg, 620002, Russia

**Abstract**

Experimental discovery that compressed sulphur hydride exhibits superconducting transition temperature $T_c$=203 K (Drozdov *et al* 2015 *Nature* **525** 73) sparked intensive studies of superconducting hydrides. However, this discovery was not a straight forward experimental examination of theoretically predicted phase, instead it was nearly five-decade long experimental quest for superconductivity in highly-compressed matters, which varied from pure elements (hydrogen, oxygen, sulphur, lithium), cuprates, and hydrides (SiH$_4$, YH$_3$, and AlH$_3$), to semiconductors and ionic salts. One of these salts was cesium iodide, CsI, which converts into metallic state at $P$=115 GPa and at $P$=180 GPa this compound exhibits the onset of the superconducting transition temperature $T_c$~2 K (Eremets *et al* 1998 *Science* **281** 1333). Detailed first principles calculations (Xu *et al* 2009 *Phys Rev B* **79** 144110) showed that within Eliashberg theory of superconductivity, the CsI should exhibits $T_c$=0.03 K at pressure $P$=180 GPa, which is by two orders of magnitude lower than the observed value. In attempt to understand the nature of this discrepancy, here we analyzed temperature dependent resistance in compressed CsI and found that this compound is perfect Fermi liquid metal which exhibits extremely high ratio of the Debye energy to the Fermi energy, $\frac{\hbar\omega_D}{k_B T_F} \cong 17$. This implies that direct utilization of the Eliashberg theory is incorrect for this compound, because the theory valid for $\frac{\hbar\omega_D}{k_B T_F} \ll 1$. We also showed that highly-compressed CsI exhibits the ratio of $\frac{T_c}{T_F} = 0.04 - 0.07$ and it falls in unconventional superconductors band in the Uemura plot.

# Fermi-liquid nonadiabatic highly-compressed cesium iodide superconductor

## I. Introduction

Since the superconducting transition at 203 K was observed in highly-compressed sulphur hydride by Drozdov *et al* [1], nearly a dozen of superconducting hydrogen-based phases have been discovered [2-18]. The report by Drozdov *et al* [1] triumphed nearly five-decade journey in the terra-incognita of hydrogen-rich and highly-compressed matter. In this journey, superconductors family was significantly extended and the superconducting transition was experimentally observed in many non-superconducting (at ambient conditions) elements/compounds. At the same time, the transition was not observed in materials for which the first principles calculations (FPC) and the Eliashberg's theory of electron-phonon mediated superconductivity predicted high critical temperature, $T_c$. We can mention AlH$_3$ [19,20] as an outstanding case of this class of materials. However, more often, the superconducting transition was observed, but predicted $T_c$ significantly exceeds experimentally observed. The most notable case for this class of materials is compressed SiH$_4$ for which Feng *et al* [21] calculated the Debye temperature of $T_\theta = 3,500 - 4,000\ K$ and $T_c \cong 165\ K$ for compressed SiH$_4$, while experiment performed by Eremets *et al* [22] showed $T_c = 7 - 17\ K$.

More intriguingly, there are several highly-compressed compounds in which experimentally observed $T_c$ significantly exceeds the calculated one. The most famous case of these highly-pressurized compounds is sulphur hydride, for which Li *et al* [23] initially predicted $T_c \cong 80\ K$. However, Drozdov *et al* [1] reported that experimentally observed transition temperature is significantly higher, $T_c \cong 200\ K$, and the latter value is in a good agreement with theoretical calculations reported by Duan *et al* [24].

Another material from this category is highly-compressed CsI, for which detailed first principles calculations performed by Xu *et al* [25] predicted $T_c = 0.03\ K$ at pressure of $P =$ $180\ GPa$, while experimental value reported by Eremets *et al* [26] is $T_c \cong 2\ K$ (at $P = 180\ GPa$).

In this work, we answer a question, why highly-compressed CsI exhibits nearly one order of magnitude higher $T_c$ in comparison with predicted value by FPC and the Eliashberg's theory of electron-phonon mediated superconductivity.

**II. Electron-phonon coupling constant and Debye temperature of CsI at $P = 206\ GPa$**

Cesium iodide is isoelectronic with noble gas solid xenon (i.e. Cs$^+$ and I$^-$ ions in ionic salt have closed xenon-like electronic shells, and the short-range interaction between Cs$^+$ and I$^-$ ions in uncompressed salt and two Xe atoms in solid xenon is identical). The main difference between Xe and CsI is the very strong Coulomb interaction in cesium iodide in comparison with xenon [27,28]. The strength of the Coulomb interaction decreases on compression, and at $P \cong 110\ GPa$ CsI is metallized [26,28].

In the theory of electron-phonon mediated superconductivity [29,30], the phonon spectrum is one of the primary properties that determines the superconducting transition temperature, $T_c$, and, thus, this is a great interest to determine main characteristic parameter of this spectrum, i.e. the Debye temperature, $T_\theta$. To do this, we fitted temperature dependent resistance, $R(T)$, curve for CsI compressed at $P \cong 206\ GPa$ (reported by Eremets *et al* [26] in their Figure 3B) to the Bloch-Grüneisen (BG) equation [31-34]:

$$R(T) = R_0 + A \times \left(\frac{T}{T_\theta}\right)^5 \times \int_0^{\frac{T_\theta}{T}} \frac{x^5}{(e^x-1)(1-e^{-x})} \cdot dx \qquad (1)$$

where, $R_0$ is the residual resistance at $T \to 0\ K$, and the second term describes the electron-phonon scattering, where $A$ and $T_\theta$ are free fitting parameters. Eq. 1 was applied to deduce the Debye temperature in many highly-compressed superconductors, for instance, in black phosphorus [35], boron [38], sulphur [37], lithium [36], $\zeta$-phase of O$_2$ [36], SnS [38], GeAs [35], $SiH_4$ [35], $H_3S$ [35,37,39], $D_3S$ [35,37], $LaH_{10}$ [33,35], $LaD_{10}$ [35,37], *C2/m*-SnH$_{12}$ [40], $Th_4H_{15}$ [41], $ThH_9$ [41], $ThH_{10}$ [41], $YD_6$ [41], metallic hydrogen phase-III [41], and $(La, Nd)H_{10}$ [42].

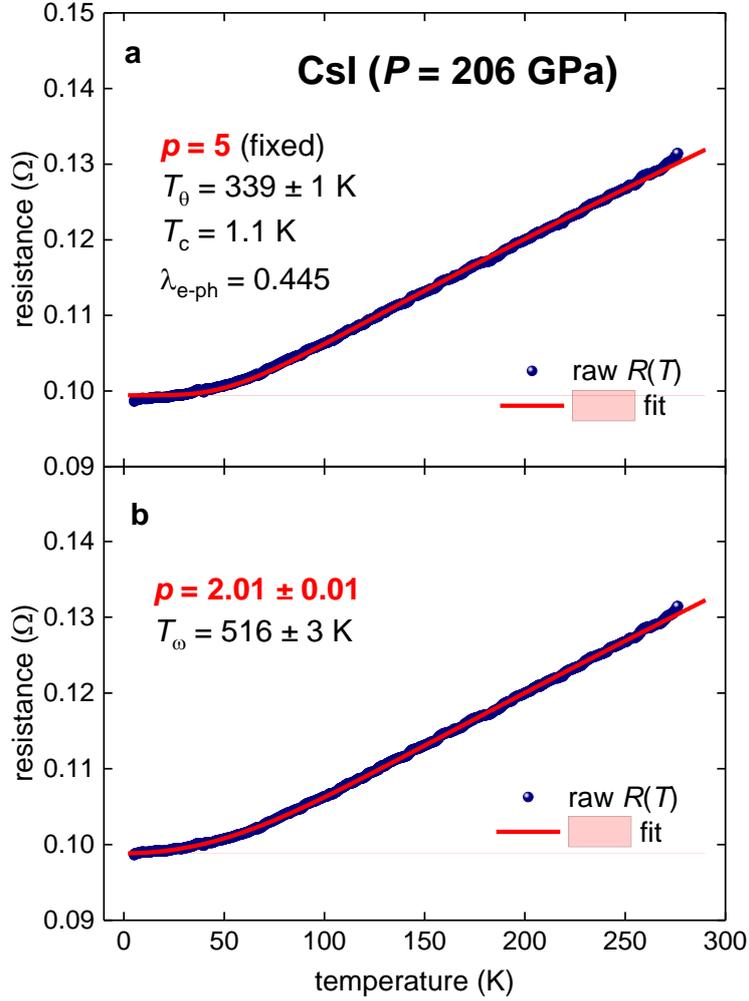

**Figure 1.** Temperature dependent resistance data, $R(T)$, in highly-compressed cesium iodide ($P = 206\ GPa$) and data fits to Eq. 1 (panel a) and Eq. 5 (panel b). Raw $R(T)$ data is from Ref. 26. (a) $p = 5$, deduced $T_\theta = 339 \pm 1\ K$, $R_0 = 0.0995\ \Omega$ fit quality is 0.9993; (b) deduced $p = 2.01 \pm 0.01$, $T_\omega = 516 \pm 3\ K$, $R_0 = 0.0988\ \Omega$, fit quality is 0.9998. 95% confidence bands are narrower than the fitting curves width.

From derived $T_\theta = 339 \pm 1\ K$ and measured $T_c \cong 1.1\ K$ [26] (Fig. 5 [26]), one can calculate the electron-phonon coupling constant, $\lambda_{e-ph}$, as the root of advanced McMillan equation [35]:

$$T_c = \left(\frac{1}{1.45}\right) \times T_\theta \times e^{-\left(\frac{1.04(1+\lambda_{e-ph})}{\lambda_{e-ph} - \mu^*(1+0.62\lambda_{e-ph})}\right)} \times f_1 \times f_2^* \qquad (2)$$

where

$$f_1 = \left(1 + \left(\frac{\lambda_{e-ph}}{2.46(1+3.8\mu^*)}\right)^{3/2}\right)^{1/3} \tag{3}$$

$$f_2^* = 1 + (0.0241 - 0.0735 \times \mu^*) \times \lambda_{e-ph}^2. \tag{4}$$

where $\mu^*$ is the Coulomb pseudopotential, it can be assumed $\mu^* = 0.13$ [35-42]. In the result, $\lambda_{e-ph} = 0.445$, which is close to $\lambda_{e-ph} = 0.43$ for aluminium [43]. It should be also noted that deduced $T_\theta = 339 \pm 1\ K$ for CsI is not much different from $T_\theta = 394 - 428\ K$ [44,45] for aluminium.

One can make a comparison of the $T_\theta = 339\ K$ and $\lambda_{e-ph} = 0.445$ values derived from experiment ($P = 216\ GPa$) with values computed by first principles calculations [25]. Xu *et al* [25] reported $\lambda_{e-ph} = 0.262$ ($P = 180\ GPa$) and $\lambda_{e-ph} = 0.257$ ($P = 216\ GPa$), and both these values are significantly lower than the one deduced from experiment herein. Xu *et al* [25] also calculated logarithmic phonon frequency $\frac{\hbar}{k_B}\omega_{log} = 285\ K$ ($P = 180\ GPa$), and $\frac{\hbar}{k_B}\omega_{log} = 314\ K$ ($P = 216\ GPa$). By its definition, $\omega_{log}$ is close, but not exact equals, to the Debye frequency, $T_\theta = \frac{\hbar}{k_B}\omega_D$, and this is what one can see for these values in highly-compressed CsI.

By utilizing the Allen-Dynes equation [46,47] and $\mu^* = 0.10$, Xu *et al* [25] calculated $T_c = 0.03\ K$ ($P = 180\ GPa$) and $T_c = 0.025\ K$ ($P = 216\ GPa$). Both calculated $T_c$ (and this was acknowledge by the authors of Ref. 25) are by about two orders of magnitude lower than experimental value. To explain this discrepancy, Xu *et al* [25] hypothesized that the Allen-Dynes equation [46,47] describes the single band superconductors and because first principles calculations showed that the crystalline structure of CsI under pressure is anisotropic, than the Allen-Dynes equation [46,47] cannot be accurate averaging technique to estimate $T_c$. Our explanation for the discrepancy is based on different idea which arose from more advanced analysis of temperature dependent resistance curve described below.

### III. Perfect Fermi liquid conductor CsI at $P = 206\ GPa$

Despite the fit of $R(T)$ curve to the BG equation (Eq. 1) has a high quality (Fig. 1,a), more advance analysis is based on an approach when the power-law exponent in Eq. 1 is a free-fitting parameter [48-50]:

$$R(T) = R_0 + A \times \left(\frac{T}{T_\omega}\right)^p \times \int_0^{\frac{T_\omega}{T}} \frac{x^p}{(e^x - 1)(1 - e^{-x})} \cdot dx \qquad (5)$$

In this approach, $T_\omega$ is not any longer the Debye temperature, however, this temperature represents characteristic energy scalar for the charge carrier interaction in the conductor.

It should be mentioned, that for some materials, like ReBe$_{22}$ [51,52] and (ScZrNb)$_{0.65}$[RhPd]$_{0.35}$ [50,53], the power-law exponent is indistinguishable from 5, which implies that these materials are pure electron-phonon mediated superconductors. However, for majority of highly-compressed superconductors, including ε-Fe phase [54], the power-law exponent, $p$, is varied between $1.80 \leq p \leq 3.3$ [50,52].

It should be mentioned that $p = 2.0$ implies that charge carriers in the conductor obey perfect Landau's Fermi liquid phenomenology [55]. The fit of $R(T)$ curve in CsI ($P = 206\ GPa$) to Eq. 5 is shown in Fig. 1,b, where deduced $p$ is indistinguishable from $p = 2.0$. This means that highly-compressed CsI at ($P = 206\ GPa$) is perfect Fermi liquid metal.

### IV. Compressed CsI ($P = 206\ GPa$) in the Uemura plot

One of the widely accepted way to classify the superconducting state in the material is to position the material in the Uemura plot, i.e. in the plot where X-axis is the Fermi temperature, $T_F$, and Y-axis is the transition temperature, $T_c$ [56,57]. If the upper critical field measurements performed, then the Fermi temperature for the material can be calculated by equation [36]:

$$T_F = \frac{\varepsilon_F}{k_B} = \frac{\pi^2}{8 \cdot k_B} \times (1 + \lambda_{e-ph}) \times \xi^2(0) \times \left(\alpha \frac{k_B T_c}{\hbar}\right)^2, \qquad (6)$$

where $\varepsilon_F$ is the Fermi energy, $k_B$ is the Boltzmann constant, $\alpha = \frac{2 \cdot \Delta(0)}{k_B \cdot T_c}$, and $\Delta(0)$ is the amplitude of the ground state energy gap, and $\hbar = h/2\pi$ is the reduced Planck constant. Based on a very large database on electron-phonon mediated superconductors [43], one can expect that CsI ($P = 206\ GPa$) which exhibites $\lambda_{e-ph} = 0.445$ should have $\alpha = \frac{2 \cdot \Delta(0)}{k_B \cdot T_c}$ not much difference from 3.53, and this value we used in our calculations. Thus, to estimate $T_F$ we estimated the ground state coherence length $\xi(0)$ by the fit of $B_{c2}(T)$ data reported by Eremets *et al* [26] (in their Figure 5) by utilizing 50% normal state resistance criterion to define $B_{c2}$. The fit to the simplest equation of the Werthamer-Helfand-Hohenberg theory [58,59]:

$$B_{c2}(0) = \frac{\phi_0}{2\pi\xi^2(0)} = -0.697 \times T_c \times \left(\frac{dB_{c2}(T)}{dT}\right)\bigg|_{T \sim T_c}, \qquad (7)$$

is shown in Fig. 2, from which $\xi(0) = 26 \pm 3\ nm$ was estimated.

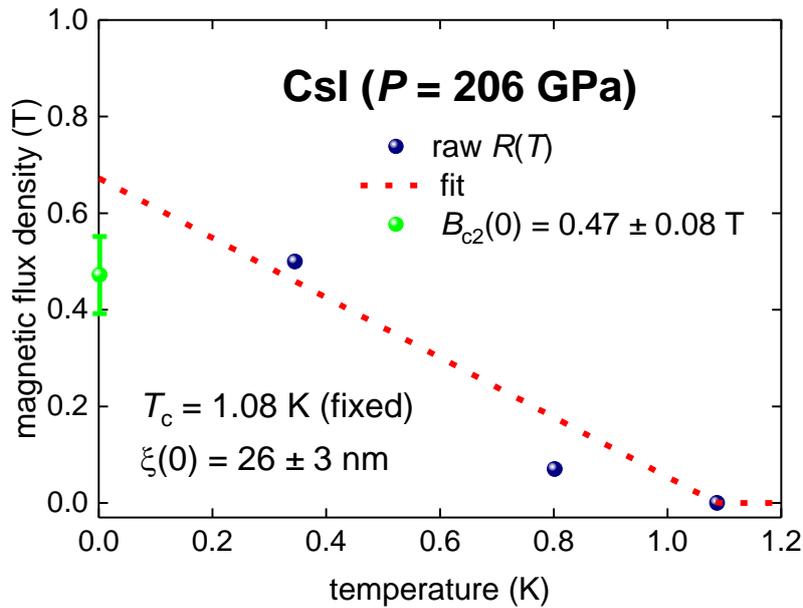

**Figure 2.** Superconducting upper critical field, $B(T)$, data (blue) for compressed CsI at pressure $P = 206$ GPa (data is from Ref. 26) and fit to WHH model [58,59] (Eq. 7) for which $T_c$ was fixed it is experimental value of 1.08 K; fit quality is $R = 0.90$.

From all determined/estimated values, one can calculate $T_F = 20 \pm 4\ K$ and the ratio of $\frac{T_c}{T_F}$, which is varying within a range:

$$0.04 \lesssim \frac{T_c}{T_F} \lesssim 0.07, \qquad (8)$$

In the result, CsI ($P = 206\ GPa$) falls into unconventional superconductors band in the Uemura plot (Fig. 3).

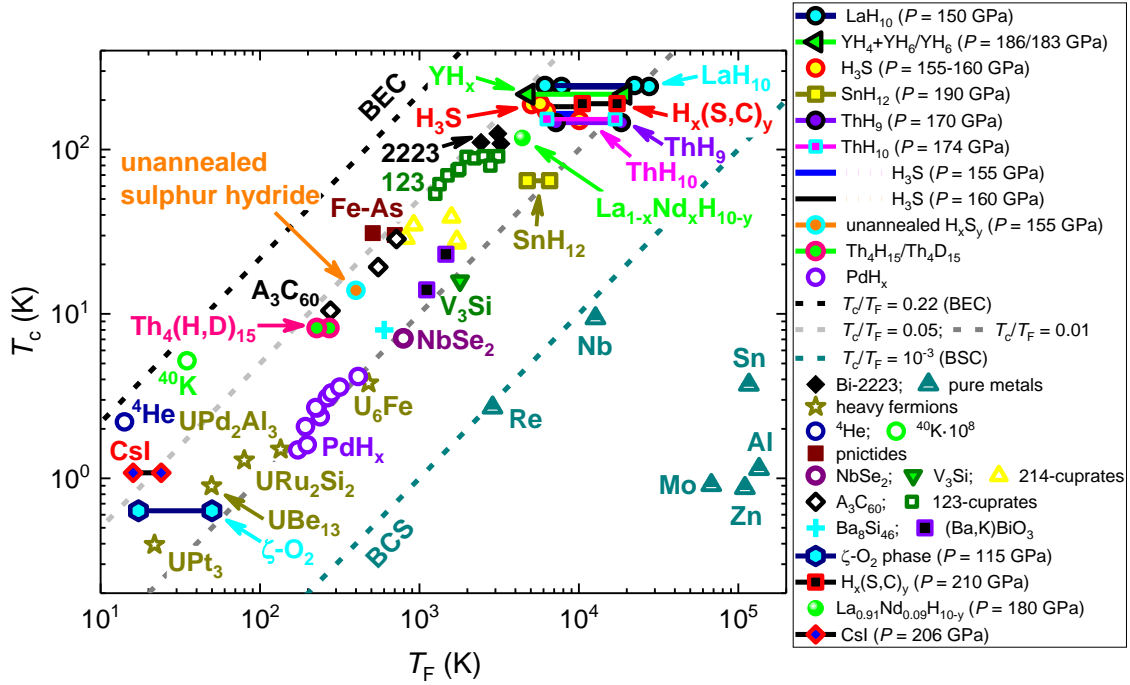

**Figure 3.** Uemura plot ($T_c$ vs $T_F$), where the CsI ($P = 206\ GPa$) is shown together with other superconducting families. References on original data ($T_c$ vs $T_F$) can be found in Ref. 40.

### V. Nonadiabatic superconductivity in CsI ($P = 206\ GPa$)

It is interesting to note that our analysis showed that CsI is remarkably prominent nonadiabatic superconductor, because this material has relatively "very fast" phonons and "very slow" charge carriers, which can be demonstrated by the ratio of:

$$\left.\frac{\hbar\omega_D}{k_B T_F}\right|_{P=206\ GPa} = \frac{T_\theta}{T_F} = 17 \pm 4 \qquad (9)$$

This kind of superconductors were first theoretically considered by Pietronero and co-workers nearly three decades ago [60-64].

This kind of superconductors is fundamentally different from traditionally considered within electron-phonon theory of superconductivity [29,30,65], which considered materials with relatively "very slow" phonons and "very fast" charge carriers. The strength of coupling does not alter remarkable difference of all these materials from CsI, because, for instance, lead and niobium [44] have the ratio of:

$$\left.\frac{\hbar\omega_D}{k_B T_F}\right|_{Pb} = \frac{T_\theta = 88\,K}{T_F = 110000\,K} = 8 \times 10^{-4} \quad (10)$$

$$\left.\frac{\hbar\omega_D}{k_B T_F}\right|_{Nb} = \frac{T_\theta = 265\,K}{T_F = 61800\,K} = 4 \times 10^{-3} \quad (11)$$

Eq. 9 shows that standard Migdal-Eliashberg theory of the electron-phonon mediated superconductivity is inapplicable for highly-compressed CsI and, thus, this is our explanation for the discrepancy between the superconducting transition temperature predicted by first-principles calculations [25] (following standard Allen-Dynes methodology [46,47]) and the observed in experiment [26].

## 4. Conclusion

In conclusion, in this paper we analysed $R(T,B)$ data for highly-compressed cesium iodide and found that this material is nearly perfect Fermi liquid conductor in the normal state. We also showed that CsI ($P = 206$ GPa) has remarkably high ratio of the phonon energy to the Fermi energy, $\frac{\hbar\omega_D}{k_B T_F} \cong 17 \pm 4$ and that this superconductor falls to unconventional superconductors band in the Uemura plot.